\documentclass[prb,superscriptaddress,aps, bibliography, twocolumn, reprint, showpacs, footnoteinbib]{revtex4-2}

\usepackage{graphicx}
\usepackage{amssymb}   % for math
\usepackage{amsfonts}
\usepackage{amsmath}
\usepackage{dsfont}
\usepackage{natbib}
\usepackage{dcolumn}   % needed for some tables
\usepackage{bm}        % for math
\usepackage[mathscr]{eucal}
\usepackage[dvipsnames]{xcolor}
\usepackage[colorlinks, linkcolor=OrangeRed,citecolor=RoyalBlue,urlcolor=NavyBlue]{hyperref}
\usepackage[all]{hypcap} % let hyperlinks correctly point to figures rather than their captions;
\usepackage{mathtools}
\usepackage{dsfont}

\definecolor{myblue}{rgb}{0,0,0.75}

\usepackage{graphicx}
\usepackage{amsmath,amssymb,bm}
\usepackage{enumerate}
\usepackage{braket}        % Dirac notation
\usepackage{physics}

\graphicspath{ {./img/} }

\usepackage{subcaption}
\usepackage{float}

\begin{document}

\title{Fate of Algebraic Many-Body Localization under driving}

\author{Heiko Burau}
\affiliation{Max Planck Institute for the Physics of Complex Systems, 01187 Dresden, Germany}
\author{Markus Heyl}
\affiliation{Max Planck Institute for the Physics of Complex Systems, 01187 Dresden, Germany}
\author{Giuseppe De Tomasi}
\affiliation{T.C.M. Group, Cavendish Laboratory, JJ Thomson Avenue, Cambridge CB3 0HE, United Kingdom}

\begin{abstract}
In this work we investigate the stability of an algebraically localized phase subject to periodic driving. First, we focus on a non-interacting model exhibiting algebraically localized single-particle modes. For this model we find numerically that the algebraically localized phase is stable under driving, meaning that the system remains localized at arbitrary frequencies. We support this result with analytical considerations using simple renormalization group arguments. Second, we inspect the case in which short-range interactions are added. By studying both, the eigenstates properties of the Floquet Hamiltonian and the out-of-equilibrium dynamics in the interacting model, we provide evidence that ergodicity is restored at any driving frequencies. In particular, we observe that for the accessible system sizes localization sets in at driving frequency that are comparable with the many-body bandwidth and thus it might be only transient, suggesting that the system might thermalize in the thermodynamic limit.      

\end{abstract}
\maketitle

\section{Introduction}
Understanding the breakdown of ergodicity in the quantum realm is an active and fast growing front of research, motivated by recent developments of controllable quantum simulations which allows one to access out-of-equilibrium dynamics \cite{Review_Bloch_2008, Gross995}. In particular, many-body localization (MBL) describes the paradigm of ergodicity breaking in quantum phases of matter, generalizing the concept of Anderson localization to the interacting case \cite{Abanin_review_2019, Rahul_2015, ALET2018498, BASKO20061126}.

An MBL phase is described by an emergent form of integrability, i.e., the existence of a complete set of quasi-local integrals of motion (LIOMs) \cite{Serbyn_con_2013, Huse_phen_2014, Imbrie2016, Ros2015}. These LIOMs are adiabatically connected to the integrals of motion of the non-interacting model and thus are typically exponentially localized.
As a consequence, transport is absent, some memory of the local structure of its initial state is retained during the quantum evolution and the interactions between the LIOMs allow a slow information propagation through the system \cite{Bar_2012, Prosen_08, Abanin_2013}.  
The MBL phase should be distinguished from an ergodic/thermal one, in which eigenstates are chaotic and are expected to obey the eigenstate thermalization hypothesis (ETH) \cite{Deutsch_91, Sre_94, Rigol2008, Luca_2016}. 

Several works have studied the stability of the MBL phase in the case of interacting one-dimensional short-range systems subject to strong disorder \cite{Prosen_08, Pal_2010, Jonas_2014, Bera_2015, Luitz_2015,DeTo_2017}. Recently, the possible existence of MBL has been shown in systems in which its single-particles are power-law localized, dubbed \textit{algebraic} MBL \cite{de2019algebraic, Bot_2019,de2019algebraic}. Algebraic MBL generalizes the paradigm of power-law localization to the interacting case and therefore its LIOMs develop algebraically decaying tails. An immediate consequence of having power-law localized LIOMs is that the bipartite entanglement entropy after a quantum quench has an unbounded algebraic growth in time \cite{de2019algebraic, Deng_2020, Naini_2019}, unlike the case with exponentially localized single-particles for which the entanglement propagates logarithmically slowly \cite{Bar_2012, Abanin_2013}. 
\begin{figure}[t!]
    \includegraphics[width=1.1\columnwidth]{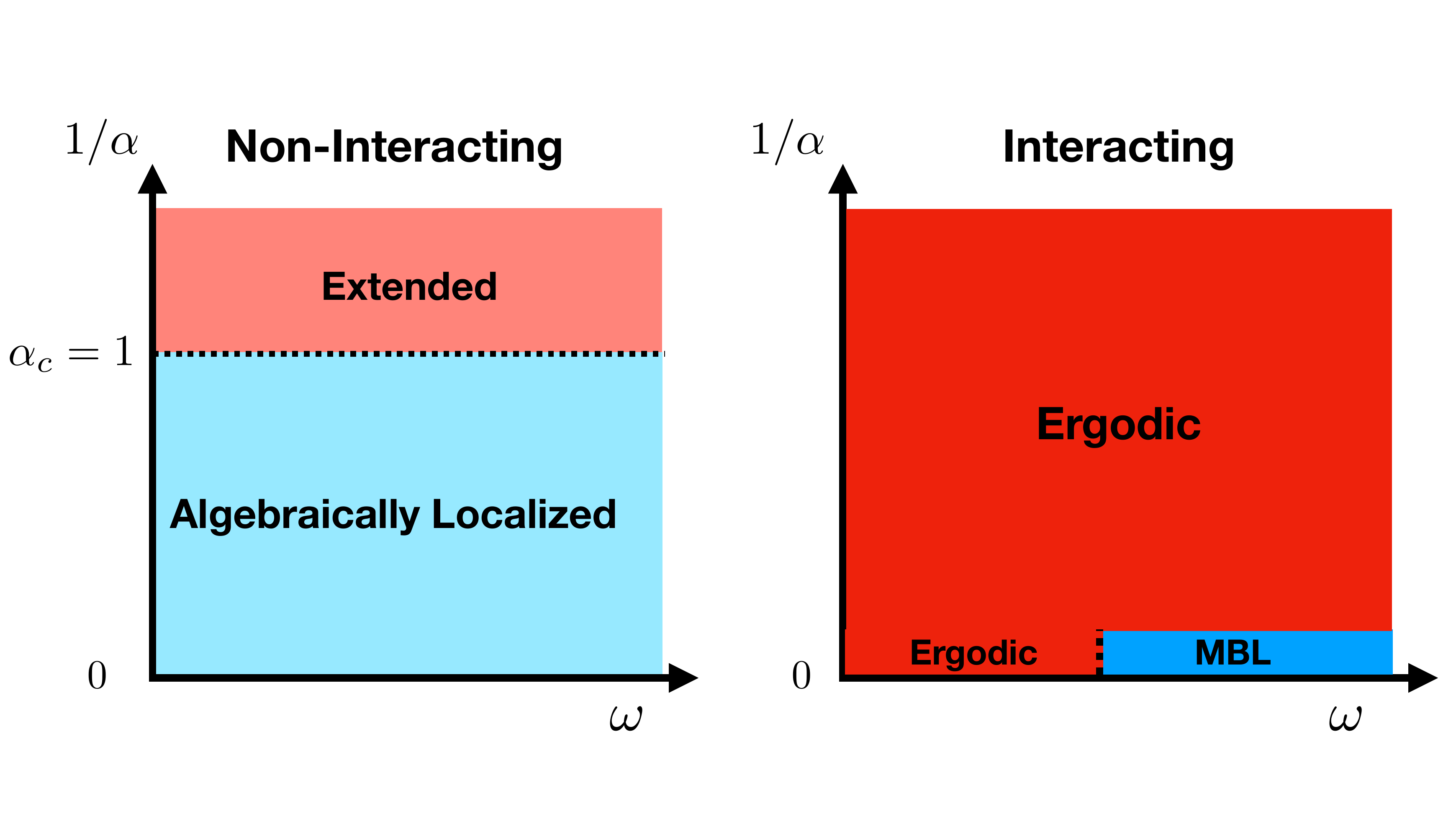}
	\caption{Schematic phase diagram of an algebraically localized phase under driving. $\alpha$ is the rate of decay of the hopping amplitudes and $\omega$ is the frequency of the driving.
	The non-interacting localized phase is robust under driving (left). Instead ergodicity is restored at any driving frequency for the interacting case (right). In the limiting case $\alpha\rightarrow \infty$ (short-range), the system undergoes an MBL transition at a finite driving frequency. 
	}
	\label{fig:picture}
\end{figure}
Furthermore, an increasing amount of efforts has been devoted to study the fate of MBL under periodic driving \cite{lazarides2015fate,ponte2015many, Wilczek_2012, Sacha_2015, Else_2016, Yao_2017, Khemani_2016, Abanin_2016, Bordia2017}. In Refs. \onlinecite{lazarides2015fate,ponte2015many} it was shown that at high frequency an MBL phase evades the fate of thermalization remaining localized, while at sufficiently low frequency the system is delocalized. 
As a consequence of the robustness of an MBL phase under driving, it has been possible to explore novel non-equilibrium quantum phases of matter, i.e., discrete time-crystals which are characterized by a spontaneous breaking of discrete time translation symmetry \cite{Wilczek_2012, Sacha_2015, Else_2016, Yao_2017, Khemani_2016}.

The aim of this work is to investigate the stability of an algebraic MBL under driving. In particular, we periodically drive (Floquet) a system of spinless fermions with long-range hopping known to have algebraically localized single-particle wavefunctions ($|\psi(x)| \sim 1/|x|^\alpha$) \cite{Deng_2018, Nosov_2019, Levitov_1989, Mirlin_1993,Levitov_90,Seli_96}.

First, we study the stability of a non-interacting algebraically localized phase by numerically inspecting several probes to distinguish a localized from an extended phase. We conclude that for the non-interacting case the localized phase does not break down under periodically driving. We support this claim by analytical considerations based on strong-disorder renormalization group techniques. Next, we focus on the interacting case. 
By studying both the eigenstates properties of the Floquet Hamiltonian and the out-of-equilibrium dynamics, we provide numerical evidence that the algebraic MBL phase breaks down under driving. We surrogate this conjecture with simple analytical considerations.  Thus, ergodicity is restored at any finite drive's frequency in the thermodynamic limit. This result might imply the non-existence of time-crystals with algebraically localized LIOMs. Figure~\ref{fig:picture} summarizes schematically the phase diagram of an (non-)interacting algebraically localized phase under driving. 

The rest of the work is organized as follows. In Sec.~\ref{sec:model} we introduce both, the non-interacting and the interacting model and we discuss the different probes that have been used to detect localization/delocalization. Section~\ref{Sec:Non-int} is devoted to probe the stability of a non-interacting case. We support these findings using a renormalization group approach, which is explained in Sec.~\ref{Sec:RG}. 
Finally, in Sec. \ref{Sec:Algebraic_MBL} we study numerically the stability of an algebraic MBL phase subject to driving. 
%Several probes, ranging from entanglement entropy to level spacing statistics have been used to demonstrate the instability of an algebraic MBL under driving. 

\section{Model $\&$ Methods}\label{sec:model}

We consider a periodically driven system of spinless fermions on an one-dimensional lattice with long-range hopping
\begin{align}\label{eq:hamiltonian} 
H(t) =  \sum_{ i \neq j } \qty(J_{ij}  c_i^\dagger  c_j + h.c.) &+ \qty(1 + \delta(t) ) \sum_i h_i  n_i \nonumber \\ &+ V \sum_i  n_i  n_{i+1},
\end{align}
where $J_{ij} = J_{ji} =  \mu_{ij} / \sqrt{1 + |i - j|^{2\alpha}}$ describes a disordered long-range hopping involving all pairs of sites, with an algebraically decaying amplitude w.r.t. their distance according to an exponent $\alpha > 0$ \cite{tran2019locality, kravtsov2015random, Deng_2018, Nosov_2019, Levitov_1989, Mirlin_1993,Levitov_90,Seli_96}. 
Quenched hopping disorder is given by $\{\mu_{ij}\}$, which is uniformly distributed between $[-1, 1]$, and on-site disorder is added by $\{h_i\}$ which are random fields uniformly distributed between $[-W,W]$. $V$ is the interaction strength and $L$ is the number of sites and we indicate with $N = L/2$ (half-filling) the number of particles.

We choose a binary driving protocol of period $T$, i.e., $\delta(t) = -\delta$ for $0 < t < T/2$ and $\delta(t) = +\delta$ for $T/2 < t < T$ and $\delta(t + nT) = \delta(t), n \in \mathbb{Z}$.
As an energy reference we define the hopping unit $J = \sqrt{\frac{1}{2L}\sum_{i\ne j} \langle {J_{ij}^2 \rangle}}$, where $\langle \cdot \rangle$ indicates the average over disorder and by $\omega_D = 2\pi/T$ we denote the driving frequency in the following.

We now briefly summarize the dynamical phase diagram of the non-driven Hamiltonian $H(\delta = 0)$ in Eq.~(\ref{eq:hamiltonian}).
The non-interacting case without driving ($V,\delta = 0$), referred as power-law banded random matrix (PLBRM), is known to have a metal-insulator transition at $\alpha = 1$ from an extended ($\alpha < 1$) to an algebraically localized phase ($\alpha > 1$) \cite{mirlin1996transition, quito2016localization, khatami2012quantum}. Thus, for $\alpha >1$ the single-particle orbitals are algebraically localized meaning that they possess power-law decaying tails around their centers of localization $i_c$, $\psi (i) \sim |i-i_c|^{-\alpha}$.  In this work, we are interested in the robustness of this localized phase under driving, therefore we mostly consider $\alpha > 1$. 
%
%As a step towards our main result, we will first study the robustness of this type of localization against external, periodic driving in the non-interacting limit.
%
%Based on exact numerically data and a dynamical extension of a flow-equation ansatz, we will show that the localized phase of the PLBRM is utterly robust against driving at arbitrary frequencies.
%
%Furthermore, we find a universal scaling-law that preserves the critical point to delocalization at $\alpha = 1$ in the thermodynamic limit, which remains valid for arbitrary driving frequencies.

%Let us now go back for a moment to the non-driven case.
%
As interactions are turned on ($V > 0$), the non-driven system goes through an MBL transition as a function of the disorder strength $W$  \cite{de2019algebraic,Rigol_Khatami_2012}. For $V=1$ and $\alpha = 3$, the critical point has been estimated to be $W_c\approx 3$, ergodic for $W<W_c$ and localized for $W>W_c$. The MBL phase of $H$ in Eq.~(\ref{eq:hamiltonian}) is described by a complete set of LIOMs $\{\tau_i\}_{i}^L$ \footnote{$[H,\tau_i] =0$ $\forall i$.} which have algebraically decaying tails, i.e, $\text{Tr}[ \tau_i A_j] \sim 1/|i-j|^{\alpha_1}$, where $A_j$ is a generic local observable  with support at site $j$ \cite{de2019algebraic}. 

It is important to point out, that the question whether MBL is stable in higher dimensions or for long-range Hamiltonians, is still under debate  \cite{Thiery_2018,Gapa_2019,Insta_2017, Wahl2019, De_Roeck_2017, kennes2018manybody, The_veniaut_2020, Altman_2019,  chertkov2021numerical,  Francesca_2021, Thomas_18, Tang2021, Kevin2021}. Recent experiments in fairly large system sizes and for relevant time-scales, have shown evidence of an MBL phase for both, two-dimensional systems and long-range Hamiltonians  \cite{Choi1547, Smith2016, Xu_2018, Guo_2020}. However, in quantum-avalanche theory the proposed mechanism for delocalization invokes the existence of rare entropic thermal bubbles, which will have effect at very long time scales and large system-sizes  \cite{Thiery_2018,Gapa_2019,Insta_2017, DeTo2020,Kos_2019}.
%Thus, to what extend the time-scales reached in experimental set-ups and the system sizes simulated, are long and large enough to claim the existence of a genuine MBL phase remain unresolved.  

We will numerically show that the algebraic MBL phase is vastly \textit{unstable} against external driving. This fragility appears to be independent of the driving frequency, although partially accompanied with finite-size effects depending on the domain of frequency. 
As a result, we observe a qualitative difference with respect to the behavior from short-ranged, or a "conventional" MBL phase under driving, in which the LIOMs are exponentially localized. For the latter, it is believed that
a system-size independent critical driving frequency $\omega_c$ exists, above which the system remains localized \cite{lazarides2015fate, ponte2015many, Abanin_2016}.

Throughout this manuscript, numerical results were obtained using free-fermions techniques for the non-interacting case and exact diagonalization for the interacting one. 
In order to establish the existence of a localized phase, we use a variety of  probes, each giving insights to the dynamical phase structure from different facets. In particular, for each quantity we average over at least $10^2$ random configurations and over energy spectrum. 

In the remainder of this section we describe all utilized probes in detail.

\subsection{Level spacing statistics}

The distribution of level spacing has been found to be an useful probe to distinguish an ergodic phase from an MBL one \cite{Luca_2016, Oga_2007, Atas_2013}. In an ergodic phase at infinite temperature, the level spacing is expected to be distributed as the one of a random matrix belonging to the same universality class of the Hamiltonian, namely the Gaussian Orthogonal Ensemble (GOE). However, in an MBL phase, due to the existence of LIOMs, energy levels are weakly correlated and the level spacings are Poisson distributed. 

Specifically, the degree of proximity to a random matrix ensemble can be extracted from the distribution of level spacing through the so called $r$-value, or gap-ratio parameter, which is defined as 
\begin{equation}
r = \langle \min(\delta_n / \delta_{n+1}, \delta_{n+1} / \delta_n ) \rangle, 
\end{equation}
where $\delta_n$ are level spacing of two consecutive level energies and $\langle \cdot \rangle$ refers to the disorder and spectrum average \cite{Oga_2007, Atas_2013}. The $r$-value in the ergodic phase is $r_{\text{GOE}} \approx 0.531$, while it is $r_{\text{Poisson}} = 2\log{2}-1$ in a localized one.

For the periodically driven system in Eq.~(\ref{eq:hamiltonian}), we consider the quasi-energies of the Floquet Hamiltonian $H_{\text{F}}$ defined by
\begin{equation}\label{eq:H_eff}
    e^{-i H_{\text{F}} T} = U(T) = \mathcal{T} e^{-i \int_0^T \dd{t'} H(t') },
\end{equation} 
which serves as generator for the discrete translation in time of period $T$ \footnote{$\mathcal{T}$ denotes the time-ordered product.}.
%
%The $r$-value is calculated numerically by taking the matrix-logarithm of the diagonalized matrix of $U(T)$ from which $H_{eff}$ and thus its eigenenergies are straightforwardly obtained.
%
%In this manuscript, the physical meaning of quasi-energies is a bit subtle: for the interacting case they refer to the many-body quasi-energies at half filling, because this is the symmetry sector we want to focus on, whereas for the non-interacting case they refer to the single-particle quasi-energies.
	
\subsection{Energy propagation}
In driven systems the energy is not conserved during time evolution and at any discrete time step an extensive amount of energy is pumped up in the system. Thus, in an ergodic phase the long-time asymptotic steady state is expected to be a structureless infinite temperature state. 
However, in an MBL phase the expected heating is effectively prevented at high driving frequencies.  This can be understood from the emergence of an extensive number of locally conserved quantities, which effectively hinder the system to exchange energy beyond a characteristic localization length  \cite{Abanin_2016, lazarides2015fate, ponte2015many}.

In order to study the energy absorbed through the quantum time evolution, we define the dimensionless parameter 
\begin{equation}
	\epsilon(\tau) = \frac{ \ev{H(\tau=0)}{\psi(\tau)} - E_{\rm min}}{E_{\rm max} - E_{\rm min}},
	\label{eq:energy}
	\end{equation}
where $\tau = t/T$ and $\ket{\psi(\tau)}$ is the time-evolved state, starting from the ground-state of $H(\tau=0)$. Here, $E_{\rm min}$ and $E_{\rm max}$ refer to the lowest and highest eigenvalues of $H(\tau=0)$, respectively. In an ergodic phase in the thermodynamic limit the asymptotic long-time value of $\epsilon(\tau \to \infty) = 1/2$ (middle of the spectrum), while in an MBL phase $\epsilon(\tau \to \infty)<1/2$.

\subsection{Entanglement entropy}

The MBL transition can be understood in terms of a rearrangement of the local structure of entanglement in the system \cite{Pal_2010, Luitz_2015, Jonas_2014}. In particular, the half-chain bipartite entanglement entropy (EE) measures the amount of entanglement through the cut and can be used, both to detect the MBL transition and to characterize the two phases. The EE ($S$) is defined through the half-partition reduced density matrix $\rho_{L/2}$, where the sites belonging to one half of the system have been traced out 
\begin{equation}
S = -\text{Tr}(\rho_{L/2} \ln \rho_{L/2}).
\label{eq:entanglement}
\end{equation}

In an ergodic phase at infinite temperature, the eigenstates behave like random vectors and therefore the EE reaches the so called Page value $S_{\rm Page} = L/2\log{2}-1/2$ \cite{Don_1993}, \footnote{It is important to point out that however also fully ergodic states can reach the Page value \cite{Ivan_multy_2020}}.

For the out-of-equilibrium dynamics, we compute the EE following a quantum quench starting from the N\'eel state $|\psi(0)\rangle = \prod_{i}^N c_{2i}^\dagger |0\rangle$. In an ergodic system driven in time, $S(\tau)$ increases up to the value of an infinite temperature thermal state, $S_{\rm Page} = L/2\log{2}-1/2$. In an MBL phase, on the other hand, $S(\tau)$ is known to saturate to an extensive value which is not thermal \cite{Bar_2012, Abanin_2013}, thus we expect $\lim_{t \to\infty} S(\tau)/S_{\rm Page} < 1$. 

Furthermore, we compute the EE for the eigenstates of the effective Hamiltonian $H_{\text{F}}$, which will have a volume-law $S \sim O(L)$ in the ergodic phase and a sub-volume scaling in the localized one $\lim_{L\rightarrow \infty} S/L \rightarrow 0$ \cite{Singh_Rajeev_2017}. 

\subsection{Inverse partition ratio}

Finally, for the non-interacting case ($V=0$), we compute the inverse partition ratio ($IPR$) \cite{Evers_2008} for the single-particle eigenfunctions of the Floquet Hamiltonian $H_{\text{F}}$, which is defined as

\begin{equation}
\label{eq:IPR}
IPR(\epsilon_\nu) = \sum_i  | \phi_\nu(i) |^4,
\end{equation}
where $i$ indicates the lattice site and $\phi_\nu(i)$ refers to the single-particle wavefunction with single-particle energy $\epsilon_\nu$. The $IPR$ separates a delocalized phase from a localized one from its scaling with $L$, $IPR\sim O(L^{-1})$ in a delocalized phase and $IPR\sim O(L^0)$ in a localized one.
	
We emphasize that in driven systems the density of states (DOS) is constant, up to finite-size fluctuations, due to the lack of energy ordering, therefore we can safely average $IPR(\epsilon_\nu)$ over all energies.  

\section{Results}

\subsection{Non-interacting case}\label{Sec:Non-int}

In this section, we inspect the eigenstate properties of the Floquet Hamiltonian $H_{\text{F}}$ for the non-interacting case. In this case, $H_{\text{F}}$ is still a quadratic Hamiltonian $H_{\text{F}} = \sum_{i,j} h^{\text{F}}_{ij} c_i^\dagger c_j$. Thus, we can simply focus on the single-particle energies $\{\epsilon_\nu\}$ and eigenstates $\{\phi_\nu\}$ of $H_{\text{F}}$.  
\begin{figure}[t!]
    \includegraphics[width=\columnwidth]{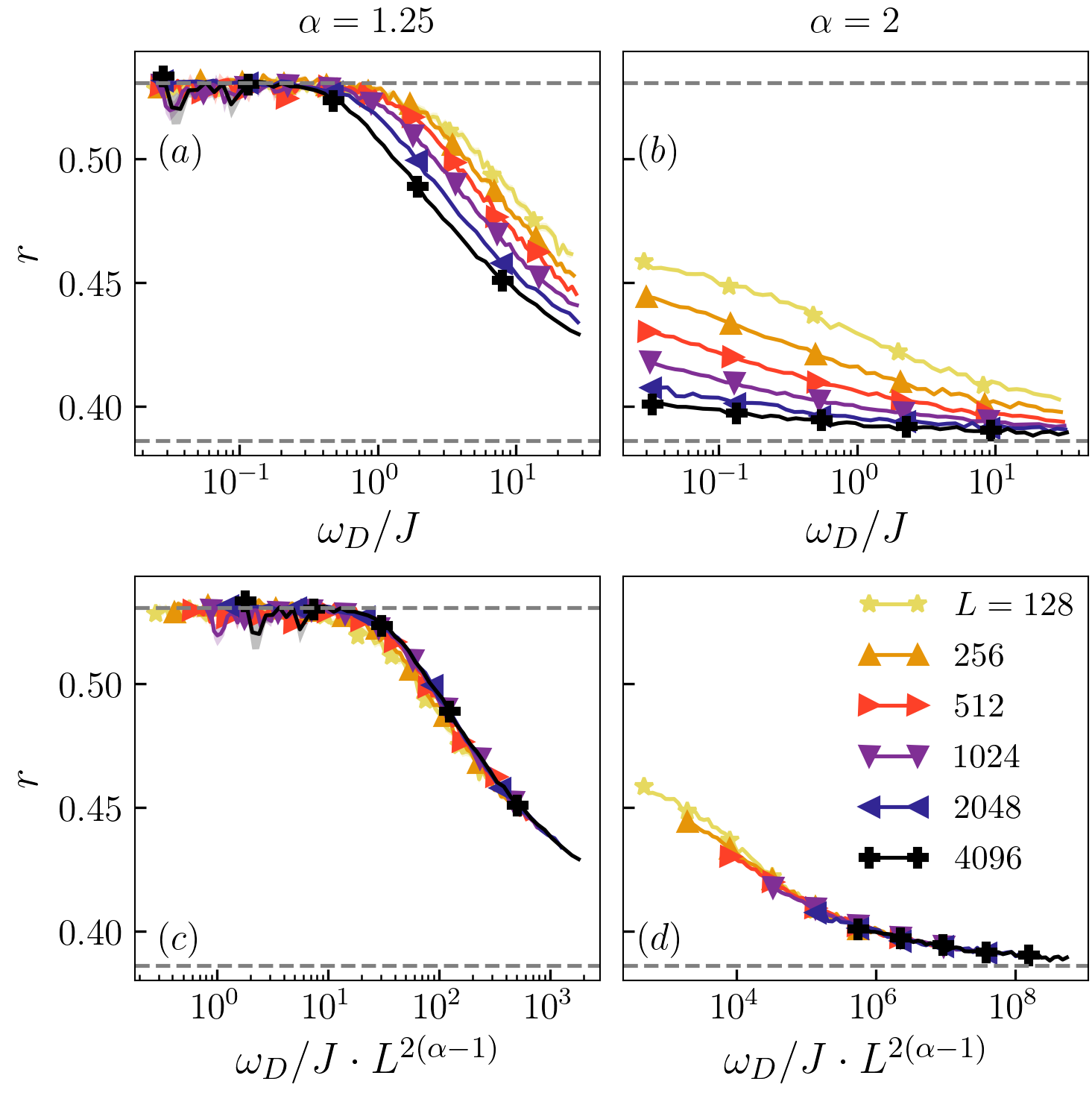}
	\caption{The $r$-statistics parameter for the quasi-eigenenergies of the non-interacting model with for $\alpha=1.25$ (a), (c) and $\alpha=2$ (b), (d) for several system sizes $L$. The panels (c), (d) show the collapse of $r$ with $L$, $r = r(\omega / J \cdot L^{2(\alpha - 1)},\delta, W)$. In all panels $W=6$ and $\delta = 0.5$. The horizontal dashed lines indicate the GOE value $\approx 0.531$ and the Poissonian one $\approx 0.386$.}
	\label{fig:collapse}
\end{figure}

\begin{figure}[ht!]
    \includegraphics[width=\columnwidth]{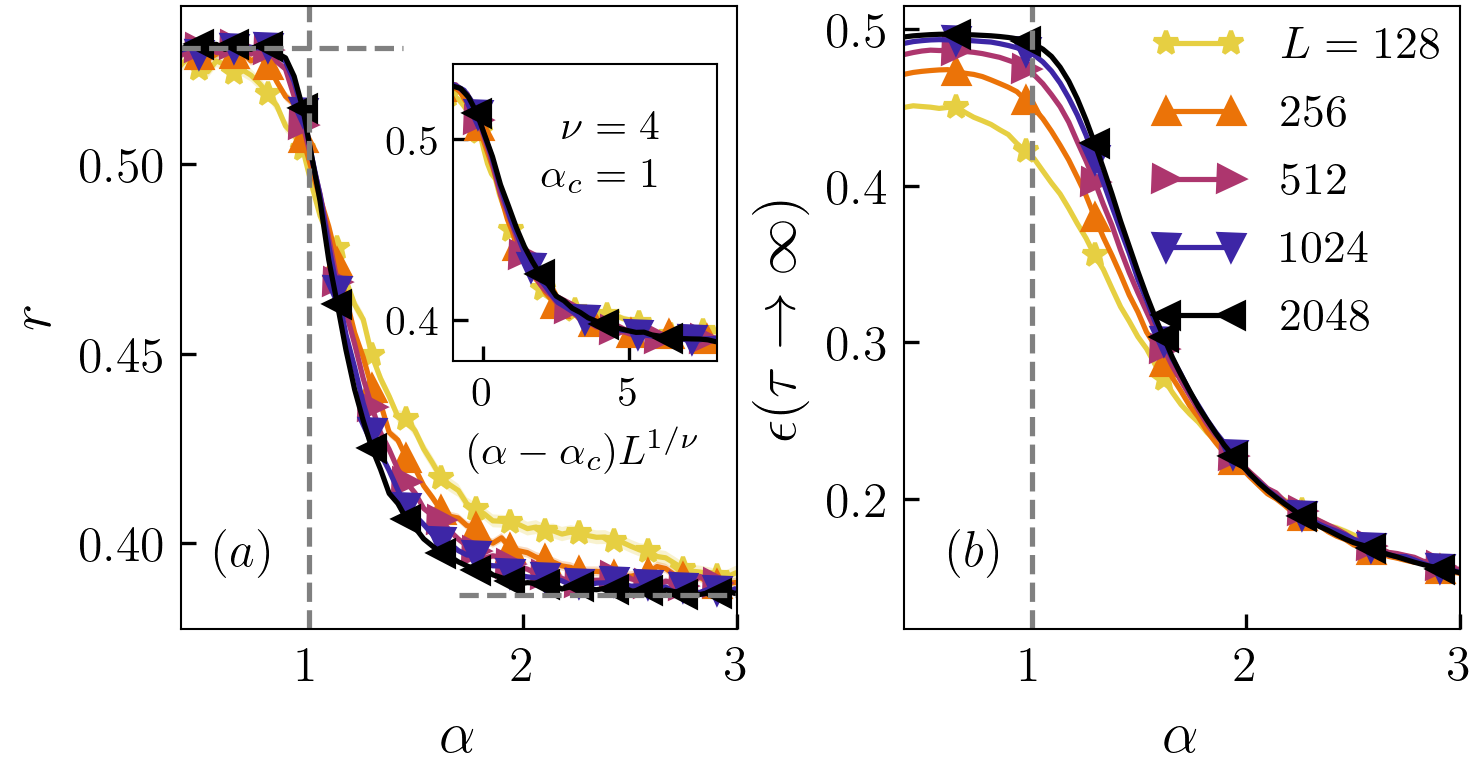}
	\caption{The $r$-statistics of quasi-eigenenergies (a) and long-time saturation value of the absorbed energy density $\epsilon(\tau)$ (b) as function of  $\alpha$ for the non-interacting case. The system is driven with a frequency  $\omega_D = J(\alpha)$ in (a) and $\omega_D = J(\alpha) / 10$ in (b), both with $W=6$ and $\delta=0.5$. The inset in panel (a) shows the finite-size scaling proving evidence of the transition at $\alpha = 1$.
	In panel (a) the horizontal dashed lines indicate the GOE value $\approx 0.531$ and the Poissonian one $\approx 0.386$. In both panels, the vertical dashed line at $\alpha = 1$ markers the delocalization-localization transition for the PLBRM}
	\label{fig:r_stats_over_alpha}
\end{figure}

We start our analysis by inspecting the level statistics of the single-particle Floquet Hamiltonian $h^{\text{F}}$. 
Fig.~\ref{fig:collapse} shows the level spacings parameter $r$ computed with $\{\epsilon_\nu\}$ as a function of the driving frequency $\omega_D = 2\pi /T$. As expected, at low frequencies ($\omega_D\ll J$) the system is more delocalized than higher ones ($\omega_D\gg J$). Indeed, at low frequencies and relatively small system sizes $L$, the $r$-value is close to the one of a random matrix ($r_{GOE} \approx 0.52$), see Fig.~\ref{fig:collapse} (a). However, an overall scaling towards localization (Poissonian statistics) with increasing $L$ is visible in Fig.~\ref{fig:collapse} (a),(b). As shown in Fig.~\ref{fig:collapse} (c), (d), we found the scaling law which describes the crossover from delocalization to localization with increasing $L$
\begin{equation}\label{eq:universal_scaling_law}
r = r(\omega / J \cdot L^{2(\alpha - 1)},\delta, W).
\end{equation}
Thus, only in the $L$-dependent frequency regime $\omega_D \ll J L^{-2(\alpha-1)}$ the single-particle wavefunction are delocalized. Importantly, the above scaling law breaks down at the metal-insulator transition ($\alpha = 1$) of the PLBRM. Indeed, for $\alpha<1$ we expect level energies to show repulsion, as visible in Fig.~\ref{fig:r_stats_over_alpha} (a), where the $r$-value is shown as a function of $\alpha$ at fixed frequency for several system sizes $L$. Finite-size scaling analyse shows that the phase diagram of the PLBRM, delocalized for $\alpha<1$ and localized for $\alpha>1$, remains unchanged under driving, see inset in Fig.~\ref{fig:r_stats_over_alpha} (a). 

\begin{figure}[ht!]
	\includegraphics[width=\columnwidth]{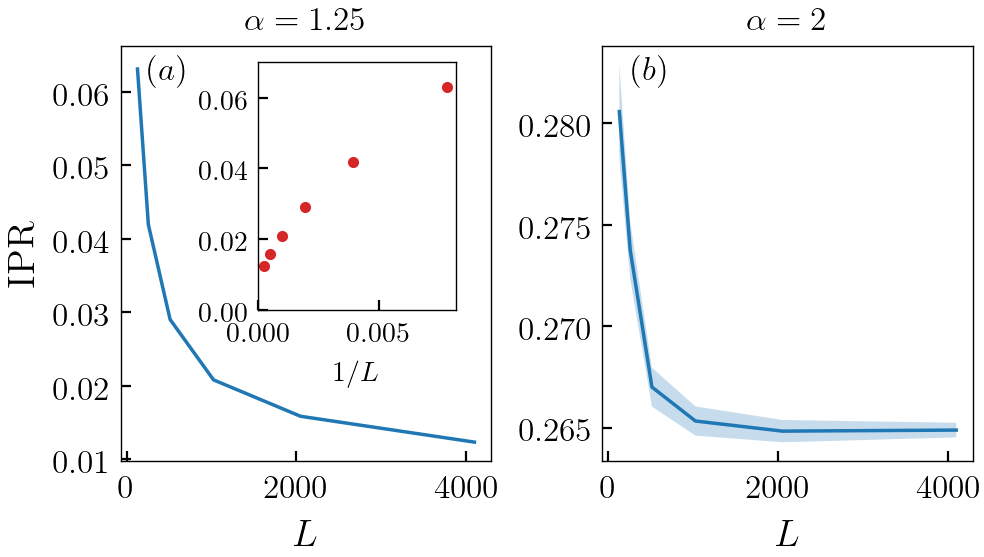}
	\caption{The inverse partition ratio ($IPR$) as function of $L$ for the single-particle wavefunctions for $\alpha=1.25$, $\alpha=2$ in panels (a) and (b), respectively.  The inset in panel (a) shows a finite-size scaling of $IPR$ with $L$, indicating that $\lim_{L\rightarrow \infty} IPR(L) \sim O(L^0)$. }
    \label{fig:IPR}
\end{figure}

Thus, we conclude that the localized phase is robust under driving, meaning that for $\alpha >1$ and at any finite frequency $\omega_D$ the spectrum of the single-particle quasi-energy remains Poissonian in the thermodynamic limit ($L\rightarrow \infty$). For the sake of completeness, we further confirm this result by inspecting the $IPR$ in Eq.~(\ref{eq:IPR}) for the single-particle wavefunctions $\{\phi_\nu\}$ of the Floquet Hamiltonian, which quantifies their degree of localisation in real space. In Fig.~\ref{fig:IPR} we present the averaged $IPR$ for two values of $\alpha$ and fixed $\omega_D / J = 0.1$. In agreement with the analysis of the $r$ level statistics, the $IPR$ converges to a finite value with $L$ ($\lim_{L\rightarrow \infty} IPR(L)>0$) and therefore the system is localized (see inset in Fig.~\ref{fig:IPR} (a) for $\alpha= 2.5$). 

We complete our analysis for the non-interacting case by investigating the behavior of two dynamical indicators, the energy transport and the information propagation quantified by $\epsilon(\tau)$ in Eq.~(\ref{eq:energy}) and $S(\tau)$ in Eq.~(\ref{eq:entanglement}), respectively. As explained in Sec.~\ref{sec:model}, the energy density $\epsilon(\tau)$ is computed starting from the ground-state of $H(\tau=0)$ and for the EE from the N\'eel state $\prod_{i}^N c_{2i}^\dagger |0\rangle$.

\begin{figure}[ht!]
    \includegraphics[width=\columnwidth]{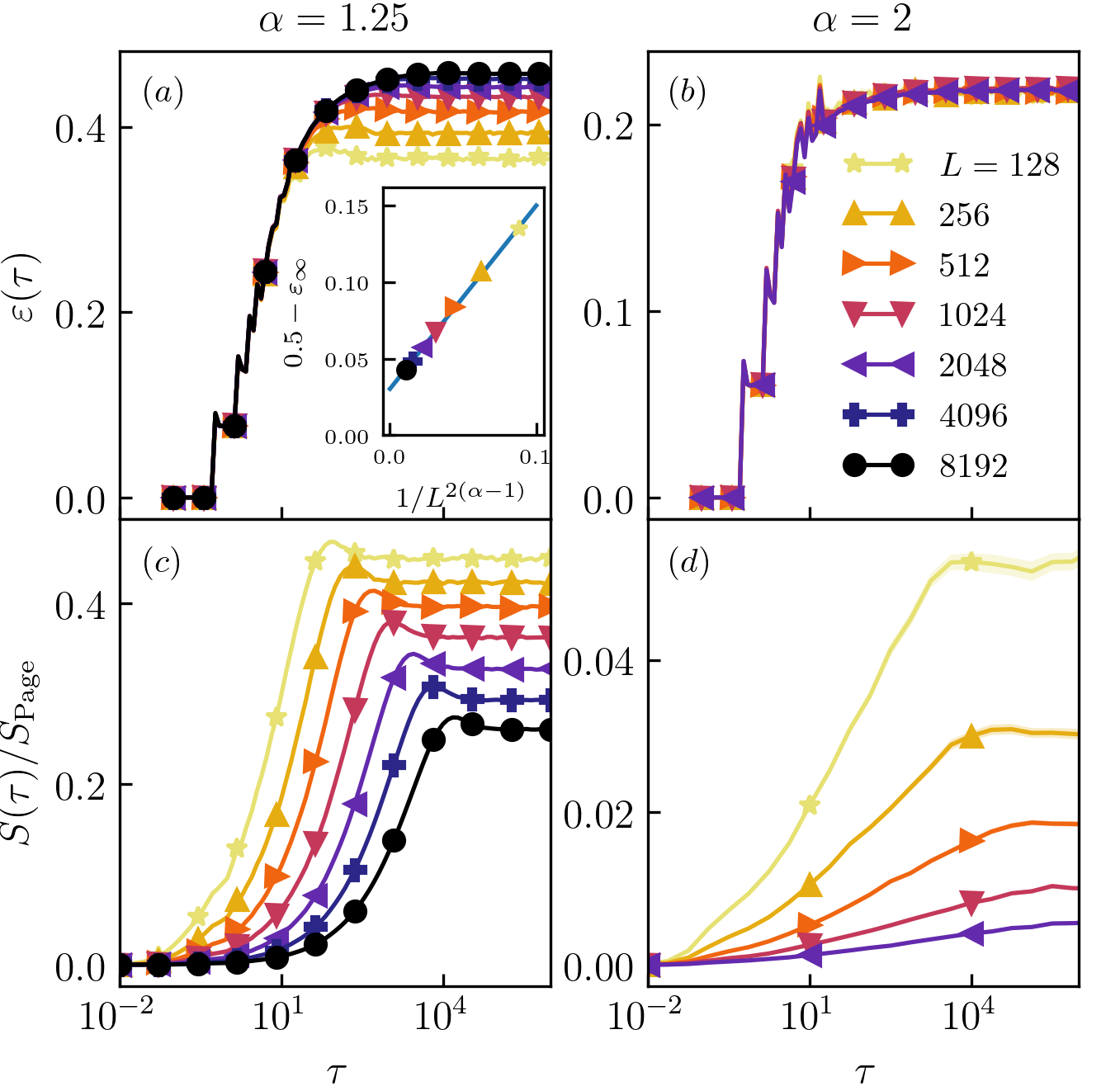}
	\caption{Absorbed energy-density $\epsilon(\tau)$ in Eq.~(\ref{eq:energy}) in (a),(b) and the half-chain bipartite entanglement entropy in (c),(d) as function of the period driving period $\tau$. The inset in (a) shows the finite-size scaling of the saturation value at asymptotic large times of $\epsilon(\tau)$. $S(\tau)$ is normalized with the value of the entanglement entropy of a random-state $S_{\rm Page} = L/2\log{2}-1/2$.}
	\label{fig:non_int_energy_and_entanglement}
\end{figure}

As expected, in the localized phase the system does not heat up to infinite temperature, meaning that $\epsilon(\tau) < 1/2$ and the saturation value of EE is sub-volume ($\lim_{L\rightarrow \infty} S(\infty)/S_{\rm Page} \rightarrow 0$) \cite{Singh_Rajeev_2017}, as one can see in Fig.~\ref{fig:non_int_energy_and_entanglement}. In particular, we study the long-time saturation value of $\epsilon(\tau)$ as function of the parameter $\alpha$. As one can see in Fig.~\ref{fig:r_stats_over_alpha} (b), $\epsilon(\infty) \approx 1/2$  for $\alpha \le 1$, meaning that in the delocalized phase energy diffuses to its maximum value. 

To summarize, by inspecting several indicators, i.e., level spacing, $IPR$, entanglement and transport properties, we have provided numerical evidence that the non-interacting localized phase remains localized once the system is driven in time.   

\subsection{Driven flow equations}\label{Sec:RG}

In the previous section, we have shown numerically that the non-interacting localized phase is robust under driving. We now present an analytical argument supporting our results. 

For this purpose, we closely follow a recent Renormalization Group (RG) approach \cite{quito2016localization, mard2014strong} which uses flow-equations \cite{kehrein2007flow} to diagonalize a static Hamiltonian.
After briefly reviewing this RG approach, we extend it to the periodic driving case, $\delta(\tau) \neq 0$, and show that the driving becomes irrelevant in the RG sense. 
Thus, we will conclude that external driving does not destroy the phase structure of the non-interacting case, in agreement with our numerical results. 

In Ref. \onlinecite{quito2016localization} Quito et al. have shown that the diagonalization of Eq.~(\ref{eq:hamiltonian}), with $V, \delta(\tau) = 0$, can essentially be decomposed into a sequence of two-body problems, i.e., a successive decimation of bonds, whereby the level-spacing statistics of the entire Hamiltonian is maintained.
This strictly holds as long as $\alpha > 1/2$ and becomes asymptotically exact in the limit $\alpha \to \infty$ \cite{quito2016localization}.
Let us consider a single RG-step, or in other words, the decimation of a bond between two lattice sites $i, j$.
Here it is convenient to define new variables, $\mathcal J = J_{ij}$ and $x = h_i - h_j$.
Now, the Wegner-Wilson flow \cite{wegner1994flow, glazek1993renormalization, glazek1994perturbative, pekker2017fixed} defines a continuous scale $\Gamma$ together with unitary transformations $U(\Gamma) = \mathcal{T}_\Gamma \exp(\int^\Gamma d\Gamma' S(\Gamma'))$ which diagonalizes the problem in the limit $\Gamma \rightarrow \infty$.
$S(\Gamma)$ is the instantaneous generator $S(\Gamma)$ and satisfies the condition $[H_0(\Gamma), S(\Gamma)] = H_1(\Gamma)$, where $H_{0} = \frac{x}{2}(n_i - n_j)$ refers to the diagonal and $H_1=\mathcal J c_i^\dagger c_j + h.c.$ to the off-diagonal part of the two-body problem (for details see Ref. \onlinecite{quito2016localization}).
Here, the explicit $\Gamma$-dependency is dropped for simplicity.
Applying $U(\Gamma)$ to the Hamiltonian, or equivalently, solving the Heisenberg equation $\partial_\Gamma H(\Gamma) = [S(\Gamma), H(\Gamma)]$, with $H(\Gamma) = H_0(\Gamma) + H_1(\Gamma)$, leads to a continuous decay of the off-diagonal matrix elements $H_1(\Gamma)$ such that $H(\Gamma)$ becomes diagonal for $\Gamma \to \infty$.
With the explicit form of $S = \mathcal J x c_i^\dagger c_j - h.c.$ the above Heisenberg equation can be straightforwardly expressed by the following coupled differential (flow-)equations,
\begin{equation}\label{eq:flow_equations}
\begin{cases}
\frac{\partial}{\partial \Gamma} \mathcal J &= -x^2 \mathcal J, \\
\frac{\partial}{\partial \Gamma} x &= 4 x \mathcal J^2 ,
\end{cases}
\end{equation}
where $\mathcal J \to 0$ for $\Gamma \to \infty$ with arbitrary initial conditions.
A characteristic RG time scale on which $\mathcal J$ decays, is given by $\tau_\Gamma = 1 / r^2$ with $r^2 = 4 \mathcal J^2 + x^2$ being an integral of motion w.r.t. $\Gamma$ in Eq.~(\ref{eq:flow_equations}).
The time scale $\tau_\Gamma$ sets an order of the RG-steps beginning with the largest $r_{max}^2$.
The fact that $r^2$ is conserved gives rise to an intuitive picture of the renormalization process: each RG-step moves one point (bond) within the $x - \mathcal J$ plane towards the $x$-axis on a circle of radius $r$.
Thereby, $x^2_{\Gamma \approx \tau_\Gamma} \to   x^2_{\Gamma = 0} + 4 \mathcal J^2_{\Gamma = 0} $ gets renormalized.
%
% In this picture, the RG transforms the initial joint distribution $P_{\Gamma = 0}(\mathcal J, x)$ into the final distribution $P_{\Gamma \to \infty}(x)$. 
%
As a side note it should be mentioned that a renormalization of adjacent bonds affected by the flow can be neglected here as the distribution $P(\mathcal J)$ behaves invariant under the full RG-flow \cite{quito2016localization}. % Of course, it's getting smaller and smaller...
The above picture of renormalization therefore connects the initial distribution $P_{\Gamma = 0}(\mathcal J, x)$ to the final one ($P_{\Gamma\to\infty}(x)$). 
The latter  is equivalent to the level-spacing distribution $P(s)$ in the limit $x\to 0$.
%
%We imply this limit in the following.

With this picture in mind, of the RG accumulating decimations on a circular shell, two phases can be immediately identified: 
(i) For an initially uniform distribution $P_{\Gamma = 0}(\mathcal J, x) \sim {\rm const}$, one obtains an extended phase, exhibiting level-repulsion, i.e., $P_{\Gamma\to\infty}(x) \sim x$. This is because the number of accumulated decimations $N_{\rm dec}$ within a shell $[r - dr, r]$ is $N_{\rm dec} \propto 2\pi r dr$ in this case \cite{quito2016localization}.
(ii) For any kind of short-ranged distribution, $P_{\Gamma = 0}(\mathcal J, x) \sim e^{-\mathcal J/\xi}$, where $\xi \ll r_{max}$ is a characteristic length scale, the RG leads to a localized phase, since $P_{\Gamma\to\infty}(x) \sim \rm{const}$ (no level-repulsion).

The initial marginal distribution $P(\mathcal J)$ in our system consists of two regimes separated by a critical $\mathcal J_c \sim \frac{1}{L^\alpha}$ \cite{quito2016localization}, set by the typical hopping amplitude at maximum distance $L$, 
\begin{equation}\label{eq:P_J}
    P_{\Gamma = 0}(\mathcal J) \sim 
    \begin{cases}  
    1 / (J^{1 + 1 / \alpha}) &, \mathcal J > \mathcal J_c \\
    L^\alpha &, \mathcal J < \mathcal J_c .
    \end{cases}
\end{equation}
As described above by means of the RG, these two regimes map to two different phases:
For $\mathcal J < \mathcal J_c$, the distribution in Eq.~(\ref{eq:P_J}) is constant, leading to level-repulsion, while for $\mathcal J > \mathcal J_c$ it is concentrated near $\mathcal J = 0$ and thus avoids level-repulsion.
At the crucial scale of the mean level spacing $\overline{\delta} \sim \frac{1}{L}$ (bandwidth can roughly be considered size-independent), the normalized critical $j_c$ defined by $j_c = \mathcal J_c / \overline{\delta} \sim \frac{1}{L^{\alpha - 1}}$ then gives a critical point at $\alpha = 1$.

\subsubsection{Dynamical extension}

In this section we derive a dynamical extension of Eqs.~(\ref{eq:flow_equations}) that accounts for a periodic driving term $\delta(t) \ne 0$ in the Hamiltonian.
As a simplification we start with a harmonic driving ansatz,
\begin{equation}
H_0(t) = \frac{x}{2} \qty( 1 + \delta e^{i \Omega t} + h.c.) ( n_i - n_j ),
\end{equation}
where $\Omega$ denotes a driving frequency and $\delta$ a dimensionless driving strength.
A time-dependent change of basis is generally given by $H'(t) = e^{-S(\tau)} H(t) e^{S(\tau)} - i  \partial_t S(\tau)$, with the Schrieffer-Wolff (SW) generator $S(\tau) = -S^\dagger(t)$.
In order to continuously dampen the off-diagonal hopping terms of $H_1$, we use the following generator
\begin{equation}
S(\tau) = \mathcal J x \qty(  c_i^\dagger  c_j + \qty(\frac{\delta}{1 - \omega}  c_j^\dagger  c_i - \frac{\delta}{1 + \omega}  c_i^\dagger  c_j ) e^{i\Omega t} ) - h.c. ,
\end{equation}
where $\omega = \Omega / x$ represents a normalized driving frequency.
For simplicity, this generator is chosen such that it does not create a time-dependence for the off-diagonal hopping terms during the flow.
%
%Otherwise the flow would become much more complicated.
%
It can be seen that the continuous change of basis with $S(\tau)$ can be expressed in closed form by the following, "driven" flow equations, where we omit the explicit $\Gamma$-dependencies
\begin{equation}\label{eq:driven_flow_equations}
\begin{cases}
 \frac{\partial}{\partial \Gamma} \mathcal J &= -x^2 \mathcal J \qty( 1 + 2 \delta^2 \frac{\omega}{1 - \omega^2} ), \\
 \frac{\partial}{\partial \Gamma} x &= 4 x \mathcal J^2, \\
 \frac{\partial}{\partial \Gamma} (x\delta) &= -4h \delta \mathcal J^2 \frac{ 1 }{ 1 - \omega^2 }.
\end{cases}
\end{equation}
A few considerations are in order.  
(i) We neglected higher-harmonic terms $\propto \delta^4$, which naturally arise from a time-dependent $S(\tau)$. 
(ii) The driving strength $\delta$ enters quadratically and therefore explains the need for a relatively large value of $\delta = 1/2$ to see significant effects from driving in our numerics.
(iii) In the infinite frequency limit $\omega\to\infty$ the flow equations just reduce to the static ones, consistent with the universal scaling law of Eq.~(\ref{eq:universal_scaling_law}).

For extraction of the phase structure we are only interested in the behavior at large RG times $\tau_\Gamma$, since they correspond to small values of $x$ of the  level-spacing.
This implies that large RG times also correspond to high values of $\omega = \Omega / x$.
% caution, thin ice: the RG-flow can lead to anything in principle.
%
The above set of equations thus can be further simplified by considering only terms up to leading order in $1/\omega$,
\begin{equation}
\begin{cases}
 \frac{\partial}{\partial \Gamma} \mathcal J &= -x^2 \mathcal J \qty( 1 - \frac{2\delta^2}{\omega} ), \\
    \frac{\partial}{\partial \Gamma} x &= 4 x \mathcal J^2, \\
    \frac{\partial}{\partial \Gamma} (x\delta) &= 0.
\end{cases}
\end{equation}
Within this approximation we find a first-order correction of the integral of motion $r^2$ to $\tilde r^2 = 4J^2 (1 + \frac{2\delta^2}{\omega}) + x^2$.
Hence, a typical level spacing from the static case gets scaled by $\frac{2\delta^2}{\omega}$.
By identifying $\tau_\Gamma \approx 1/x^2$, the driving correction becomes manifestly irrelevant in the course of the RG, $\frac{2\delta^2}{\omega} \approx \frac{1}{\sqrt{\tau_\Gamma}} \frac{2\delta^2}{\Omega}$. As a result, the non-interacting algebraically localized phase is robust against driving.

%In the following we will show how a finite $1 / \beta$ leads to a level repulsion in the two-site problem and how this result can be generalized to the $n$-site problem.
%\begin{equation}
%    r^2 = 4J^2 + h^2 +  4 \mathcal J^2 \frac{2\delta^2}{\beta} - \mathcal J^2 \frac{2\delta^2}{\beta} \sum_{k=1}^\infty (-1)^k a_k \qty( \frac{\mathcal J}{h} )^{2k},
%\end{equation}
%where $a_{k+1} = 4 (2k - 1) / (2k + 4) a_k, a_1 = 4$.

\subsection{Algebraic MBL}\label{Sec:Algebraic_MBL}

Having established that the non-interacting algebraic phase is stable under periodic driving, we now turn to the interacting case by setting $V=1$ in Eq.~(\ref{eq:hamiltonian}). With the aim to understand the robustness of an algebraic MBL phase, we fix the parameters $\alpha$ and $W$ of $H$ in Eq.~(\ref{eq:hamiltonian}) such that the non-driven system is localized  \cite{De_To_2019, Rigol_Khatami_2012, Singh_Rajeev_2017}. The numerical results have been computed using exact diagonalization techniques, thus we will be limited by relatively small systems $L\in [6,16]$. However, by studying several probes to distinguish an ergodic phase from an MBL one, even if affected by  finite-size effects, we provide numerical evidence that at any finite frequency, ergodicity is expected to be restored in the thermodynamic limit. 

\begin{figure}[t]
    \includegraphics[width=\columnwidth]{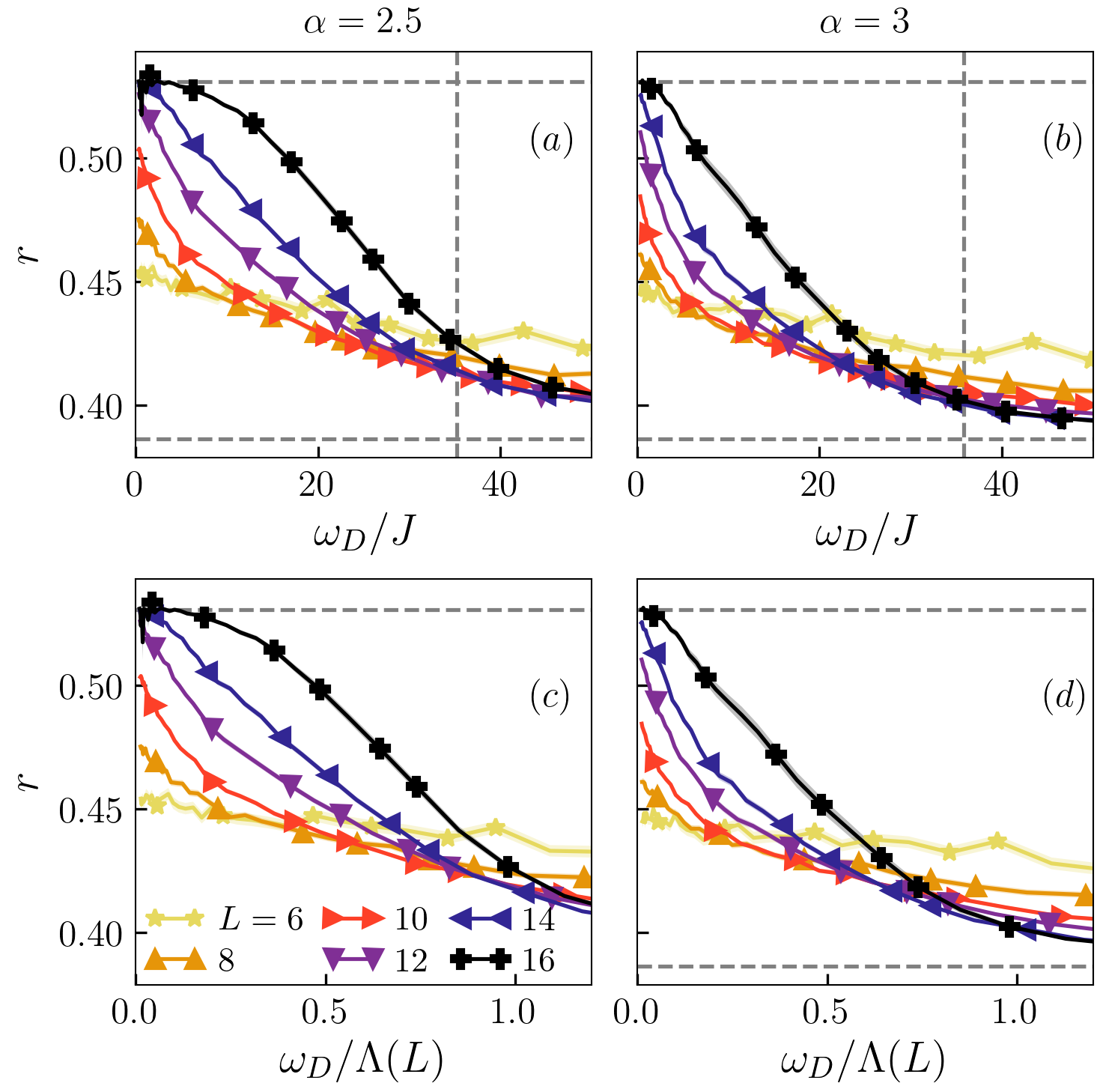}
    \caption{$r$-statistics for the interacting case ($V=1$) as a function of the driving frequency $\omega_D$, for two values of the hopping decaying-rate $\alpha = 2.5, 3$ in panel (a) and (b), respectively. The dashed horizontal lines indicate the GOE $r$-value $\approx 0.531$ and the Poissonian one $\approx 0.386$, whereas the vertical lines indicate the spectral width $\Lambda(L=16)$. In panels (c),(d), we show the $r$-value as function of $\omega_D/\Lambda(L)$. 
    In all panels $W=6$ and $\delta = 0.5$.}
    \label{fig:r_stats_int}
\end{figure}
\begin{figure}[t]
    \includegraphics[width=\columnwidth]{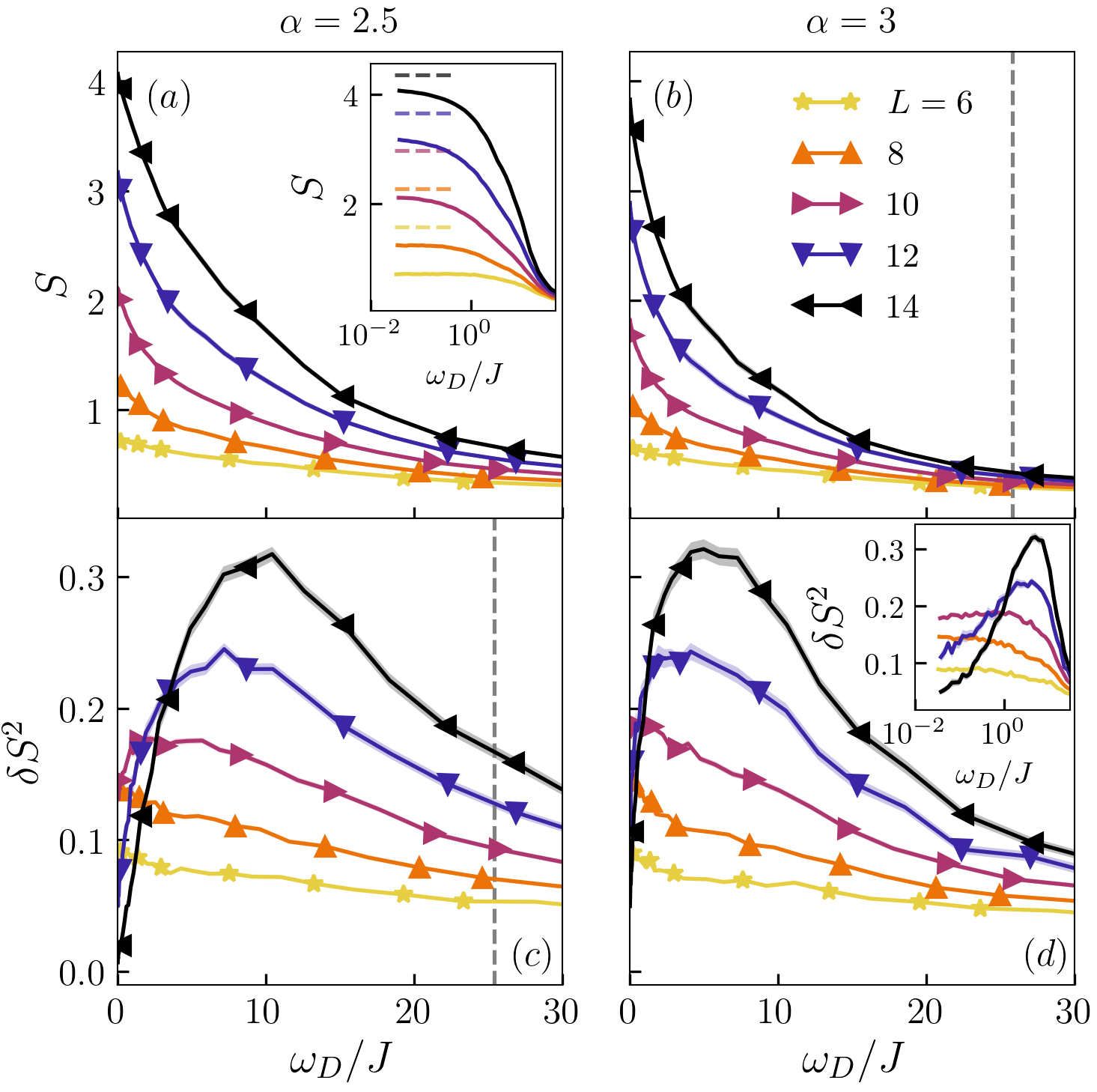}
    \caption{The averaged entanglement entropy $S$ of the Floquet Hamiltonian for $V=1$ and the variance of the entanglement entropy $\delta S^2$ in panels (a)-(b) and (c)-(d), respectively. 
    The inset in (a) shows $S$ at smaller frequencies for several $L$ with respective Page values. $\delta S^2$ at smaller frequencies for $\alpha= 3$ is shown in the inset of panel (d). The dashed vertical lines indicate the bandwidth $\Lambda(L)$.}
    \label{fig:statics_int}
\end{figure}
\begin{figure}[t]
    \includegraphics[width=\columnwidth]{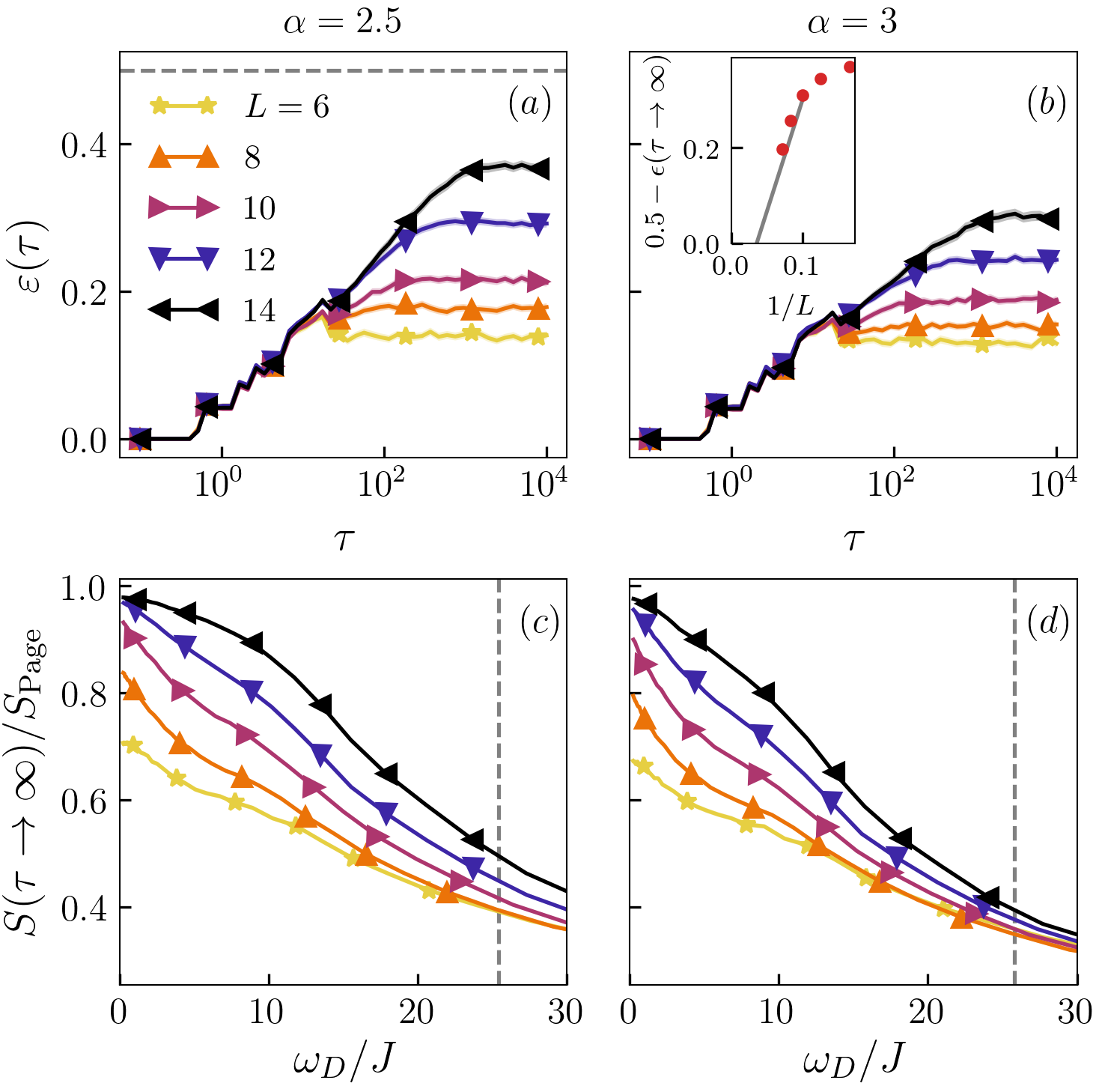}
    \caption{In panels (a)-(b) are shown the energy absorption rate $\epsilon(\tau)$ in Eq.~(\ref{eq:energy}) starting from the ground-state of the interacting model for two values of $\alpha$ for the interacting case. The inset in panel (b) shows the finite-size scaling for the long-time saturation value of $\epsilon(\tau)$.
    In panels (c)-(d) the long-time saturation point for the bipartite entanglement entropy with the initial state given by the N\'eel state ($\prod_{i}^N c^\dagger_{2i} |0\rangle$). The dashed vertical lines indicate the bandwidth $\Lambda(L)$.}
    \label{fig:dynamics_int}
\end{figure}

As well as for the non-interacting case, we start our analysis by inspecting the spectrum of the effective Hamiltonian $H_{\text{F}}$ in Eq.~\ref{eq:H_eff}. Fig.~\ref{fig:r_stats_int} shows the $r$-value as function of the driving frequency $\omega_{D}$, for several $L$. A clear trend towards delocalization ($r\approx 0.531$) with increasing $L$ is visible in Fig.~\ref{fig:r_stats_int} (a)-(b). The $r$-value change their behavior at high-frequency $\omega_D \gg J$, where a frequency-independent plateau is formed close to the Poissonian value $r\approx 0.386$. This plateau belongs to the finite size perturbative regime, in which the system is close to the non-driven case $U(T) \approx \mathbb{I} -\frac{i T}{2}(H(T/2) +H(T))$, which disappears in the thermodynamic limit. Generically, a driving localized phase remains localized at enough high frequencies. In particular, this will happen if the driving frequency $\omega_D$ exceed the many-body bandwidth $\Lambda(L)$ of the energy spectrum of the non-driven Hamiltonian \cite{lazarides2015fate}. We compute the averaged density of states of the non-driven Hamiltonian and using a Gaussian fit we extrapolate its typical bandwidth $\Lambda(L)$ (vertical dashed line in Fig.~\ref{fig:r_stats_int} for the largest $L$). 
\begin{figure}[t]
    \includegraphics[width=\columnwidth]{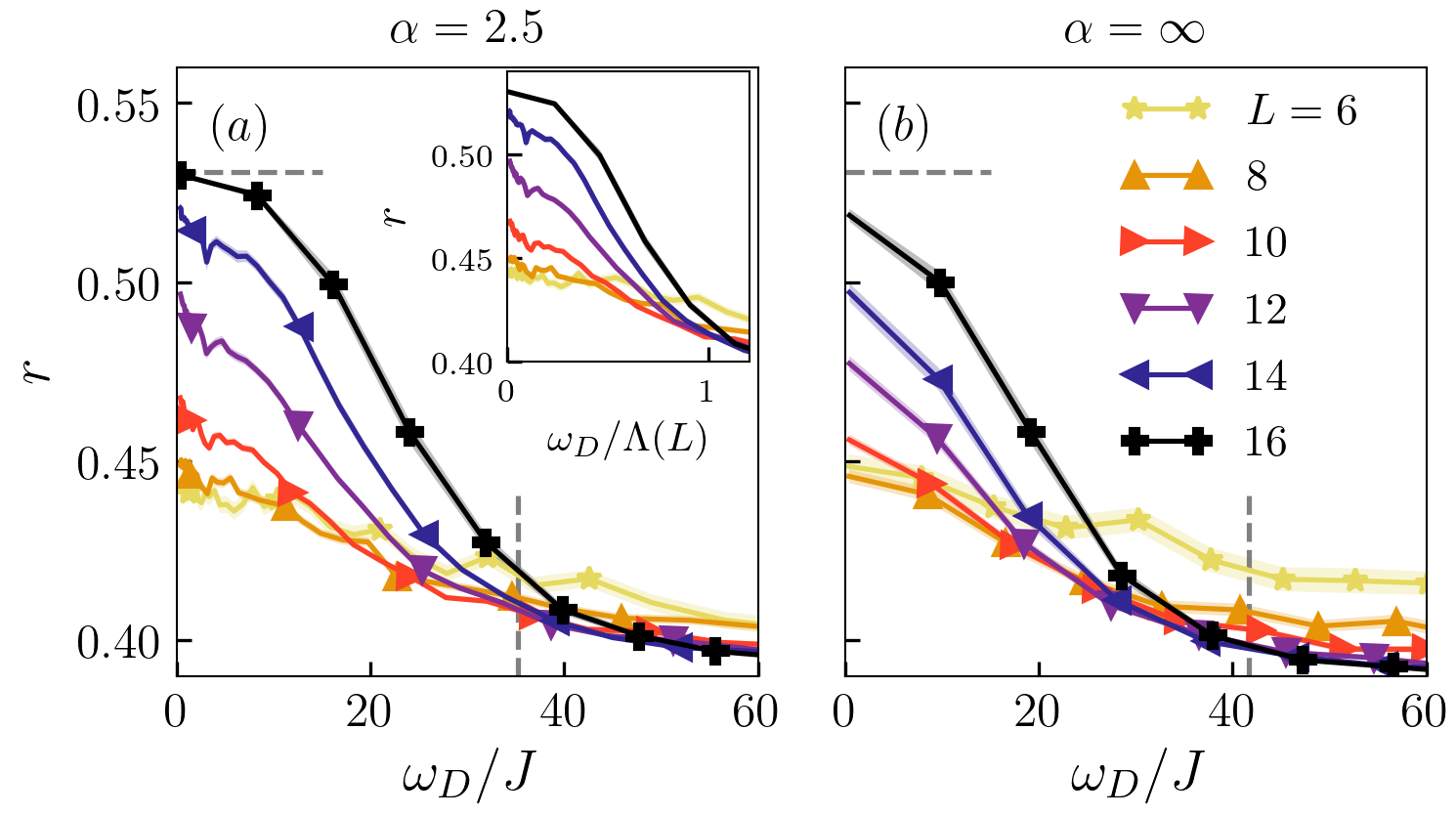}
    \caption{$r$-statistics as a function of the driving frequency $\omega_D$, for the hopping decaying rate $\alpha = 2.5$ in panel (a) and $\alpha \rightarrow \infty$ in (b). The dashed horizontal lines indicate the GOE $r$-value $\approx 0.531$ and the Poissonian one $\approx 0.386$, whereas the vertical lines indicate the spectral width ($\Lambda(L=16)$). In the inset of panel (a) is shown the $r$-value as a function of $\omega_D/\Lambda(L)$ for $\alpha = 2.5$. In both panels $W=6$ and $\delta = 0.5$.}
    \label{fig:r_stats_int_toy_model}
\end{figure}

Thus, for $\omega_D \ge \Lambda$ our numerical result cannot be used to understand the dynamical phase diagram of $H(t)$. By safely ignoring the high-frequency regime ($\omega_D \ge \Lambda$), the $r$-value of $H_{\text{F}}$ shows a clear tendency towards delocalization  (rigid spectrum), allowing us to claim that ergodicity might be restored in the thermodynamic limit. This is visible in Fig.~\ref{fig:r_stats_int} (c),(d), where $r$-value is showed as function of $\omega_D/\Lambda(L)$, providing evidence that the critical frequency $\omega_c$ for which localization sets on, scales as $\omega_c \sim \Lambda(L)\sim \sqrt{L}$. 

With the aim to further support the claim of delocalization of an algebraic MBL phase under driving, we study the entanglement properties of the eigenstates of the Floquet Hamiltonian ($H_{\text{F}}$). Fig.~\ref{fig:statics_int} shows the bipartite EE averaged over disorder and eigenstates and its fluctuations $\delta S^2$ as function of the driving frequency \cite{Jonas_2014, Luitz_2015}. The EE has a similar behavior as the $r$ level statistics parameter, meaning that low frequency EE reaches the ergodic values (see inset in Fig.~\ref{fig:statics_int} (a)) and then the curves indicate a shift to delocalization (volume law scaling) before approaching the natural cut-off $\Lambda$. Furthermore, we study fluctuations around the mean value $S$, which are quantified by the variance ($\delta S^2$) over random samples and eigenstates \cite{Jonas_2014, Luitz_2015}. At low frequencies $\delta S^2$ decays with $L$ in agreement with the ETH (see inset in Fig.~\ref{fig:statics_int} (d)). Meanwhile at moderate frequencies $\delta S^2$ develops a growing peak with $L$,  a feature which has been used as a fingerprint for the 
MBL transition \cite{Jonas_2014, Luitz_2015}. However, as one can observe, the peak shifts inexorably with system size, getting closer to $\Lambda$.   

Finally, we inspect some dynamical properties of the interacting driven system, i.e., energy and entanglement propagation. 

As for the non-interacting case, we start with the propagation of energy. Fig.~\ref{fig:dynamics_int} (a)-(b) shows the energy density $\epsilon(\tau)$ at moderate frequency $\omega_D \approx 4J$. 
For both values, $\alpha = 2.5, 3$ and for our available system sizes, the infinite temperature value ($\varepsilon = 1/2$) is not reached yet. However, in long-time limit, $\epsilon(\tau)$ tends to $\varepsilon \approx 1/2$, which it also supported by the finite-size scaling (see inset in Fig.~\ref{fig:dynamics_int}). The remaining two panels, Fig.~\ref{fig:dynamics_int} (c), (d), are dedicated to the asymptotic long-time saturation value of $S(\tau)$. As a function of the driving frequency, we compute  $S(\tau\rightarrow \infty)/S_{\rm Page}$, where $S(\tau\rightarrow \infty)$ is determined by averaging over a long-time window in which $S(\tau)$ reaches its saturation value. The numerical analysis of $S(\tau\rightarrow \infty)/S_{\rm Page}$ in Fig.~\ref{fig:dynamics_int} (c), (d), in which is visible the tendency $S(\tau\rightarrow \infty)/S_{\rm Page}\rightarrow 1$ as $L$ increases. Thus, in agreements with the results of the spectrum and eigenstates of $H_{\text{F}}$, the finite size analysis of  $S(\tau\rightarrow \infty)$ further supports the conjecture that ergodicity is restored in the thermodynamic limit. 

Unlike the case in which the LIOMs have exponentially decaying tails ($\alpha \rightarrow \infty$), the algebraic MBL phase shows a clear tendency towards delocalization at any finite-driving frequencies.  To better pin-down this fundamental difference, we study a Floquet minimal MBL model \cite{de2019algebraic, De_To_2019, Wu_2019}, which can be considered as the strong disorder limit of $H$ in Eq.~(\ref{eq:hamiltonian})
\begin{equation}
    U(T) = e^{-i H^{+} T/2} e^{-i H^{-} T/2},
\end{equation} 
with
\begin{equation}
\label{eq:toy}
     H^{\pm} = \sum_l \epsilon_l^{\pm} I_l^{\pm} + V \sum_{p,q} B_{pq}^{\pm} I_p^{\pm} I_q^{\pm},
\end{equation}
where $\{\epsilon_l^{\pm}\}$ and $\{I_l^\pm\}$ are the single-particle energies and LIOMs of the non-interacting model ($V=0$) in Eq.~(\ref{eq:hamiltonian}) with $\pm \delta$, respectively \cite{de2019algebraic, De_To_2019}. The couplings $B_{pq} = \sum_{i} \{ |\phi_p^\pm(i)|^2 |\phi_q^\pm(i+1)|^2 - \phi_p^\pm(i)\phi_p^\pm(i+1)\phi_q^\pm(i)\phi_q^\pm(i+1)  \}$ are obtained using first order perturbation theory in $H$ in the limit $V/W\ll 1$  \cite{de2019algebraic, De_To_2019}. By construction, $H^{\pm}$ is localized meaning that it has Poissonian spectrum and integrals of motion are algebraically localized for $\alpha <\infty$ and exponential as $\alpha \rightarrow \infty$. As a further consequence, for $\alpha<\infty$ the couplings $B_{pq}$ decay algebraically w.r.t. the distance between the centers of localisation of $\phi_p$ and $\phi_q$, while for $\alpha=\infty$ the decay is exponentially fast. The spatial structure of the decay of $B_{pq}$ determinants the growth in time of the EE after a global quantum quench with $H^{\pm}$. Thus, for $\alpha<\infty$ the EE has an unbounded algebraic growth, while for $\alpha=\infty$ the growth is logarithmic slow. 

Returning back to the driven case, in Fig.~\ref{fig:r_stats_int_toy_model} (a)-(b) is shown the $r$-value as function of the driving frequency $\omega_D$ computed with the eigenvalues of the Floquet Hamiltonian of the minimal model in Eq.~(\ref{eq:toy}) for $\alpha = 2.5$ (algebraically LIOMs) and $\alpha\rightarrow \infty$ (exponentially LIOMs). The crossing between curves for different system sizes, which may demarcate the existence of a putative MBL transition as function of $\omega_D$, is located in proximity of the many-body width cut-off for $\alpha = 2.5$ and at smaller frequencies for $\alpha\rightarrow \infty$. Considering  only values $\omega_D\le \Lambda$ (see Fig.~\ref{fig:r_stats_int_toy_model} (a) and its inset), and the tendency towards delocalization for the algebraic case, we may conclude that the localization will be completely destroyed in the thermodynamic limit. Indeed, the crossing between curves for different $L$ happens for frequencies $\omega_D \ge \Lambda$. However, in the limiting case of exponential LIOMs ($\alpha \rightarrow \infty$), we observe a much more robust and defined cross-over between the two regimes which occurs before the natural bound $\omega_D \sim \Lambda(L)$, and therefore it may culminate to an MBL transition at a finite frequency strength \cite{Abanin_2016, lazarides2015fate, ponte2015many}. 

Now, we provide an analytical argument based on Ref. \onlinecite{Abanin_2016} to support the delocalization of an algebraic MBL phase under periodic drive. For the sake of simplicity, consider the case where the driving terms contain only of one harmonic and the time-dependent Hamiltonian can be separated in two parts $H(t) = H_{\ell-\text{bit}}(W) + V H_1(t)$. The Hamiltonian $H_{\ell-\text{bit}}(W)$ is algebraic MBL and $H_1(t) \sim \cos(2 \pi t/T)$ is the time-depended part which allows jumps between the localized eigenstates of $H_{\ell-\text{bit}}(W)$. $W$ and $V$ are the disorder and interaction strength, respectively. We assume that the disorder is strong enough to localize the instantaneous eigenstates. If $T$ is large enough, we can estimate the probability of mixing the instantaneous eigenstates $|E_\alpha (t) \rangle$, $|E_\beta (t)\rangle$ using the Landau-Zener formula \cite{Zener:1932ws, Stueckelberg,Majorana1932} 
\begin{equation}
P_{\alpha \rightarrow \alpha} = e^{-C_{\alpha, \beta}}, \quad C_{\alpha, \beta} = \pi \frac{|M_{\alpha,\beta}|^2}{v_{\alpha,\beta}}
\end{equation}
where $M_{\alpha,\beta} = \langle E_\alpha| H_1 (t_c)|E_\beta \rangle$ is the matrix element coupling the eigenstates, the Landau–Zener velocity $v_{\alpha,\beta} = T \partial_t (E_\alpha - E_\beta)|_{t=t_c}$ and $t_c$ the time at which the crossing takes place. In the adiabatic ($C_{\alpha,\beta} \approx 0$) and diabatic ($C_{\alpha,\beta} \gg 1$)  limits the system remains localized after the crossing. In the intermediate situation  eigenstates are characterized by different values of LIOMs mixing together and starting to thermalize. 
As argued in Ref. \onlinecite{Abanin_2016}, there exist two length scales $x_1$ and $x_2$, which determine the behavior of 
$P_{\alpha \rightarrow \alpha}$. Consider two eigenstates $|E_\alpha (t) \rangle$, $|E_\beta (t)\rangle$ which differs from the values of LIOMs only in a region of size $x$. Thus, these eigenstates are the same outside a region of size $x$. There are $\sim 2^x$ of these eigenstates which typically belong to an energy shell with bandwidth $\sim W\sqrt{x}$ and the mean level spacing 
\begin{equation}
\Delta (x) = W\sqrt{x}/2^{x}.
\end{equation}
On the other hand, the typical gap at the crossing is controlled by $V H_1(t_c)$ 
\begin{equation}\delta E_{\alpha, \beta}(x)\sim V ( \langle E_\alpha |H_1|E_\alpha \rangle - \langle E_\beta |H_1| E_\beta \rangle) \sim V \sqrt{x}.
\end{equation}
In the limit $\delta E_{\alpha, \beta} \gg \Delta$ multiple crossing between eigenstates which differ only in the region of size $x$ happen. This sets the length scale $x_1$ by comparing $\delta E_{\alpha, \beta}(x_1) \sim \Delta (x_1)$
\begin{equation}
x_1 = \log_2 \frac{W}{V}.
\end{equation}
The length $x_2$ determinants the character of the transition, which is directly connected to $C_{\alpha,\beta} \propto |M_{\alpha,\beta}|^2/v_{\alpha,\beta}$.
The matrix element between two algebraically localized states can be estimated by
\begin{equation}
M_{\alpha,\beta} \sim V/x^\gamma, 
\end{equation}
which decays as a power-law in distance $x$ since the LIOMs are algebraically localized. Instead, the Landau–Zener velocity is
\begin{equation}
v_{\alpha,\beta} \sim \Delta E_{\alpha,\beta}/T\sim
V \sqrt{x}/T.
\end{equation}
Thus, the length $x_2$ is defined by $C_{\alpha,\beta}(x_2)\sim 1$ which separate the adiabatic limit from the diabatic one,
\begin{equation}
    x_2 = (VT)^{\frac{1}{2\gamma+1/2}}.
\end{equation}
In the regime $x_1/x_2\ll 1$, the system undergoes multiples crossing in an intermediate regime $C_{\alpha,\beta}\sim 1$ which implies the thermalization of the region of size $x \ll x_1 \ll x_2$. The relation $x_1/x_2\ll 1$ implies that
\begin{equation}
\label{eq:frequency_for_deloc}
    \frac{W}{V} \ll 2^{(VT)^\frac{1}{2\gamma+1/2}}.
\end{equation}
Although this simple argument (Eq.~(\ref{eq:frequency_for_deloc})) does not strictly prove the delocalization of the system at any frequency $\omega_D = \frac{2\pi}{T}$, it shows that the delocalization is parametrically favorable, meaning that it is exponentially sensitive to $\omega_D\sim 1/T$. 

%The instantaneous eigenstates of $H(t)$, $|E_{\underline{\alpha}(t)} \rangle$ are algebraically localized and thus they can be label by a set of quantum numbers ${\underline{\alpha}=(\alpha_1,...,\alpha_L)}$. 

\section{Conclusion}\label{Sec:conclusion}

In this work we have investigated the stability of an algebraically localized phase subject to periodically global driving for both, the non-interacting and the interacting case. For sufficiently strong disorder, the non-interacting algebraically localized phase is robust once short-range interactions are turned on, generalizing the paradigm of algebraic localization to the interacting case (algebraic MBL). As a consequence, in an algebraic MBL phase, ergodicity is broken and the phase is fully characterized by an extensive number of integrals of motion with algebraic decaying tails, which are believed to be adiabatically connected to the algebraic decaying integrals of motion of the non-interacting case. 

In particular, we studied a periodically driven (Floquet) spinless fermionic chain with long-range random hopping and density-density short-range interactions. The non-driven non-interacting limit is known as the power-law random banded matrix model and hosts an algebraically localized phase, meaning that its single-particle eigenstates are localized with power-law tails. 

First, we considered the non-interacting case. With a combination of analytical arguments and exact numerical simulations, we provided evidence that the algebraically localized phase is stable under driving. The spectrum of the signle-particle Floquet Hamiltonian shows Poissonian statistics and its eigenstates are localized. We supported this claim by using analytical arguments based on strong disorder renormalization group techniques.  

Second, we have tackled the interacting case by turning on short-range density-density type interactions. At strong disorder and for sufficiently large decay rate of the hopping the non-driven case is believed to be in an algebraic MBL phase. Several MBL markers have been used to numerically inspect the stability of the algebraic MBL phase under driving. After a careful analysis of the finite-size effects, we argue that ergodicity might be restored in the thermodynamic limit for any driving frequencies. For finite-size systems we showed that localization sets in for driving frequency that are comparable with the many-body bandwidth. This indicates that, the finite-size localized phase might disperse in the thermodynamic limit. Analytical arguments were made to support our numerical results by showing that delocalization appears parametrically favorable. 

Thus, unlike the non-interacting case, we showed that the algebraic MBL phase is unstable once the system is periodically driven. This result should be compared to the case in which the LIOMs are exponentially localized. For conventional MBL phases, it is believed that a finite, critical frequency exists, which separates an ergodic phase from an MBL one. To further point out this difference, in the final part of the work we inspected a toy Floquet model, in which we can tune the type of localization, from algebraic to exponential. We confirmed the existence of a critical driving frequency in a conventional MBL phase and showed the breakdown of an algebraic MBL phase under driving. 

Our work imposes strong constraints on the possible existence of a genuine MBL phase for long-range models under driving. This result could be relevant for trapped-ions experiments, where long-range interactions are present. It is important to point out that even though we do not expect the existence of an MBL phase, interesting non-ergodic behavior could persist for fairly long time scales and large systems. Thus, understanding the time scales for thermalization is an important question which is left for future research.

%
%
%After that, we will present an intuitive picture of how driving destroys the original, algebraically localized MBL phase.

%This might point to a lower rate of thermalization in the regime of fast driving but for sure one would expect a strictly constant value in case of a driven MBL phase.

\begin{acknowledgments}
We would like to thank S. Bera and I. Khaymovich for helpful discussions. We also express our gratitude to S. Woo Kim for a critical reading of the manuscript. This project has received funding from the European Research Council (ERC) under the European Union’s Horizon 2020 research and innovation program (grant agreement No. 853443), and M.H. further acknowledges support by the Deutsche Forschungs- gemeinschaft (DFG) via the Gottfried Wilhelm Leibniz Prize program.

\end{acknowledgments}

\bibliography{references}

\end{document}